\documentclass[10pt,a4paper,twoside]{article}
\usepackage{epsfig}
\usepackage{baltlat6}
\usepackage{array}
\usepackage{wrapfig}
\usepackage{longtable}
\usepackage{here}

\begin{document}
\ \
\vspace{0.5mm}
\setcounter{page}{141}
\vspace{8mm}

\titlehead{Baltic Astronomy, vol.\,18, 141--159, 2009}

\titleb{EXTINCTIONS AND DISTANCES OF DARK CLOUDS FROM\\ UGRIJHK
PHOTOMETRY OF RED CLUMP GIANTS:\\ THE NORTH AMERICA AND PELICAN
NEBULAE COMPLEX}

\begin{authorl}
\authorb{V. Strai\v{z}ys}{} and
\authorb{V. Laugalys}{}
\end{authorl}

\moveright-3.2mm
\vbox{
\begin{addressl}
\addressb{}{Institute of Theoretical Physics and Astronomy, Vilnius
University,\\  Go\v{s}tauto 12, Vilnius LT-01108, Lithuania;
straizys@itpa.lt}
\end{addressl}
}

\submitb{Received 2009 August 15; accepted 2009 August 28}

\begin{summary} A possibility of applying 2MASS {\it J, H, K}$_s$, IPHAS
{\it r, i} and MegaCam {\it u, g} photometry of red giants for
determining distances to dark clouds is investigated.  Red clump giants
with a small admixture of G5--K1 and M2--M3 stars of the giant branch
can be isolated and used in determining distances to separate clouds or
spiral arms.  Interstellar extinctions of background red giants can
be also used for mapping dust surface density in the cloud.
\end{summary}


\resthead{Extinctions and distances of dark clouds from UGRIJHK
photometry}
{V. Strai\v{z}ys, V. Laugalys}

\sectionb{1}{INTRODUCTION}

The determination of distances to dust and molecular clouds still
remains a problem.  One of the most popular methods in use is based on
the extinction vs. distance plot -- at the cloud distance, a steep rise
in extinction takes place.  The input observational data on stars,
necessary for applying the method, are the apparent magnitudes and color
indices (usually $V$ and $B$--$V$), and two-dimensional spectral types
(usually in the MK system) which are used to estimate of intrinsic color
indices and absolute magnitudes.  Magnitudes and color indices can be
taken from observations in any photometric system, for example, in the
system of the near infrared 2MASS survey ($K_s$ magnitudes and
$H$--$K_s$ color indices).  Two-dimensional spectral types of stars can
be determined either by classification of their spectra or by applying
interstellar reddening-free quantities in one of multicolor photometric
systems, for example, in the seven-color {\it Vilnius} photometric
system.

The estimation of absolute magnitudes is the weakest point of the
method.  The absolute errors of $M_V$ estimated from MK spectra can be
as large as $\pm$\,0.5 mag, and this leads to the distance errors from
--20\,\% to +26\,\%.  For supergiants, the absolute error of $M_V$ can
reach even $\pm$\,1.0 mag.  Therefore, for more accurate distance
determination a statistically significant number of stars is essential.

In the case of clouds located at distances $>$\,1 kpc, two-dimensional
classification of stars is complicated due to faintness of field stars,
located at the cloud distance or in its background, especially in the
blue spectral range.  Multicolor photometric systems, such as the {\it
Vilnius} system (Strai\v{z}ys 1992), provide a means to classify stars
using the interstellar reddening-free $Q$ parameters, and they are able
to reach much fainter and more distant stars than in the case of
spectral classification.  However, their usefulness declines when heavy
interstellar extinction is present.

In such a case, near-infrared photometry is more favorable. 2MASS
photometry in the $J$, $H$ and $K_s$ passbands (at 1.24, 1.66 and 2.16
$\mu$m) provides an effective tool for penetrating to distant sites of
the Galaxy, hidden behind dense dust clouds.  However, in the case of
heavy and variable interstellar reddening, two-dimensional
classification of stars using the infrared colors becomes impossible.
For example, the sequences of stars of different luminosities in the
two-color diagram $J$--$H$ vs.\,$H$--$K_s$ overlap even in the absence
of interstellar reddening.  The shift of stars along reddening lines
complicates the situation even more.

Fortunately, in some of the cases, even near infrared photometry may be
helpful.  With a certain degree of confidence the use of interstellar
reddening-free parameters $Q_{JHK}$ enables us to identify K-type
giants, AGB stars, OB stars and young stellar objects (YSOs) even at
extremely large interstellar reddening (see Comeron \& Pasquali 2005;
Strai\v{z}ys \& Laugalys 2007, 2008a,b,c, Strai\v{z}ys et al. 2008).
Since this identification is not always single-valued, color indices or
magnitudes, measured in other photometric passbands, could
give the missing information in some cases.

For the determination of extinction and distance from
{\it J, H, K}$_s$ photometry we need to have a type of stars which is
easily recognizable in the presence of heavy and variable interstellar
reddening as only in such case one can estimate intrinsic colors and
absolute magnitudes of individual stars.  Such stars, shown to be
extremely useful for the extinction investigation up to great distances,
are red clump giants (hereafter RCGs) -- stars in the post-helium-flash
evolutionary stage, analogues of horizontal-branch stars of Population II.
According to our previous investigation (Strai\v{z}ys \& Lazauskait\.e
2009), their mean intrinsic color indices are:  $J$--$H$ = 0.46 and
$H$--$K_s$ = 0.09.

Although spectral types of RCG stars are in the range from G8\,III to
K2\,III, their absolute magnitudes in the $K_s$ passband, $M_{K_s}$,
show a surprising uniformity.  Using the {\it Hipparcos} parallaxes,
Alves (2000) has found that RCGs show a very small dispersion of
absolute magnitudes in the $K$ passband, $M_K$ = --1.61\,$\pm$\,0.03
mag, without dependence on small differences in the temperature and
metallicity.  Later on, a similar conclusion was reached by Grocholski
\& Sarajedini (2002) who obtained for RCGs in open clusters $M_K$ =
--1.62\,$\pm$\,0.21 mag.  These stars would be perfect distance and
reddening indicators if we were able to identify them among thousands of
stars with various spectral types and reddenings.

In the paper Strai\v{z}ys \& Laugalys (2008c) we concluded that RCGs,
intermingled with cooler stars of the giant sequence, are especially
useful for the determination of slopes of interstellar reddening lines
in the $J$--$H$ vs.\,$H$--$K_s$ diagram.  In the present paper we
investigate a possibility of isolating RCGs from other K--M\,III stars
and using them for the cloud distance determination.  We apply our
method to the region near the Galactic plane, which includes the North
America and Pelican nebulae (hereafter NAP) and the dark cloud between
them.  In this region, deep photometry is available in several
broad-band systems:  the 2MASS {\it J,H,K}$_s$ system, the IPHAS $r$,$i$
system and the MegaCam $u$,$g$ system.

\newpage


\begin{figure}[!ht]
\centerline{\psfig{figure=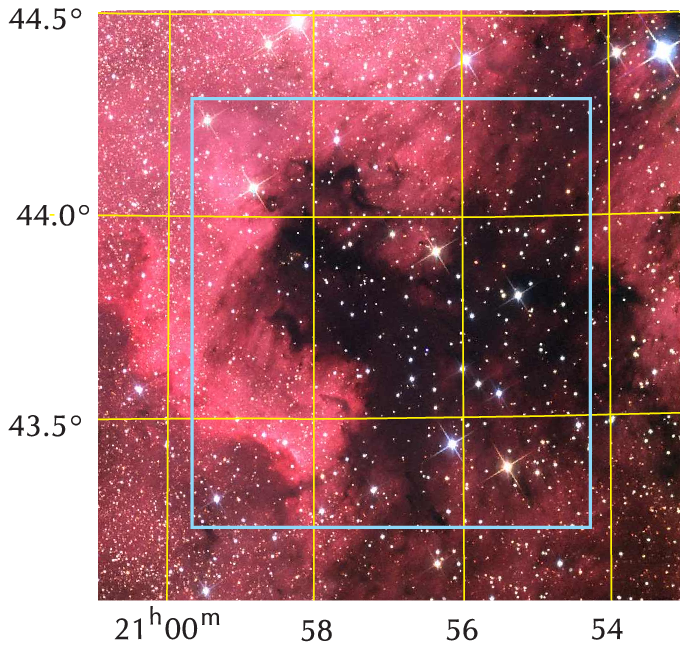,width=124mm,clip=}}
\vskip3mm
\captionb{1}{The investigated area in the North America and Pelican
nebulae complex (within the blue square).}
\end{figure}

\sectionb{2}{THE AREA IN CYGNUS}

The selected 1\degr\,$\times$\,1\degr\ area in the NAP complex with the
center at RA\,(J2000) = 20h57m, DEC\,(J2000) = +43\degr\,45\arcmin\ is
shown in Figure 1. It covers the southern part of the dark cloud
LDN\,935 (Lynds 1962) = TGU\,497 (Dobashi et al. 2005), including the
Gulf of Mexico, the `coastal' areas around the Gulf, the Florida
peninsula and the dark cloud up to the Pelican beak.  The choice of this
area was predetermined by the availability in it of deep photometry in
the $u$ and $g$ passbands, obtained with the MegaCam mosaic CCD camera
on CFHT (Gwyn 2008; MegaPipe 2009).  The $u$ and $g$ filters of the
MegaCam are similar, but not identical, to the Sloan Digital Sky Survey
(SDSS) filters.  The SDSS filter $u$ has a `red leak' at 710 nm, which
precludes from applying the $u$--$g$ index for intrinsically red and
heavily reddened stars (Covey et al. 2007).  In the $u$ filter of
MegaCam the red leak is absent.  The limiting $u$ magnitude of the
MegaCam catalog with an accuracy of $<$\,0.1 mag is about 22.5.


\begin{figure}[H]
\vbox{
\centerline{\psfig{figure=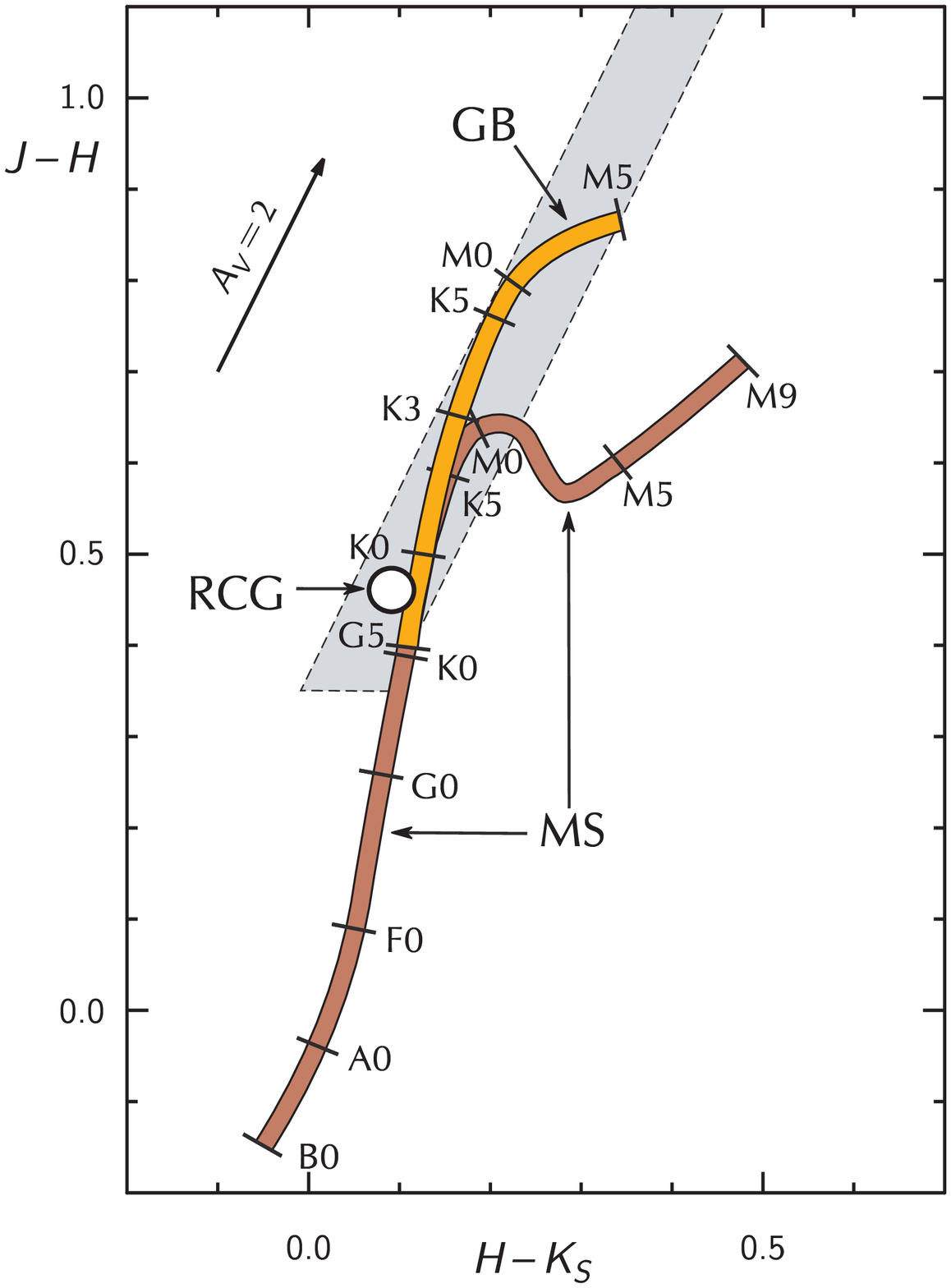,width=100mm,angle=0,clip=}}
\vspace{-.5mm}
\captionb{2}{The $J$--$H$ vs.\,$H$--$K_s$ diagram with the intrinsic
main sequence (MS, brown belt), giant branch (GB, orange belt) and the
locus of red clump giants (RCG, white circle).  Two parallel broken
lines mark the region in which RCGs with different interstellar
reddenings are located (grey strip).  The length of the reddening vector
corresponds to $A_V$ = 2 mag.}
}
\end{figure}

The $r$ and $i$ magnitudes were taken from the INT Photometric H$\alpha$
Survey (IPHAS) since the original SDSS photometry in the whole area is
not available.  This survey covers the northern Milky Way between
Galactic longitudes 30--220\degr\ and latitudes $\pm$\,5\degr\ (Drew et
al. 2005).  The objects are measured in two broad-band filters, $r'$ and
$i'$, and one narrow-band filter placed on H$\alpha$.  The $r'$ and $i'$
filters are similar to those of the SDSS
$u'$,\,$g'$,\,$r'$,\,$i'$,\,$z'$ system, their mean wavelengths are at
624 nm and 774 nm.  For simplicity, in this paper we will omit the
`prime' designations of the IPHAS magnitudes.

To define the identification criteria of RCG stars we investigated their
location in the $J$--$H$ vs.\,$H$--$K_s$ and $u$--$g$ vs.\,$r$--$i$
diagrams with respect to the sequences of luminosity V and III stars
(main-sequence stars and giants) in the presence of considerable and
variable interstellar reddening.

\sectionb{3}{RED CLUMP GIANTS IN THE J--H, H--K$_{\rm s}$ DIAGRAM}

Figure 2 shows the intrinsic sequences of luminosity V and III stars in
the 2MASS $J$--$H$ vs.\,$H$--$K_s$ diagram according to Strai\v{z}ys \&
Lazauskait\.e (2009).  The sequence of unreddened main-sequence stars of
spectral classes from A to M has been determined using field stars up to
40 pc from the Sun with available MK spectral types and stars of the
Praesepe, Pleiades and M\,39 open clusters with small dereddening
corrections.  Intrinsic colors of O and B stars have been determined by
dereddening the corresponding stars in the Ori OB1 association, the
Collinder 121 and the Pleiades clusters and field O-type stars with low
reddenings.  For determining the intrinsic colors of G5--M5 III stars,
dereddened giants from seven open clusters and field giants from the NGP
region have been used.  The approximate color indices of RCGs in the
diagram have been estimated by locating the center of the apparent
crowding of dereddened positions of the cluster giants.

It is evident that the $J$--$H$ vs.\,$H$--$K_s$ diagram does not allow
us to isolate RCGs in the presence of interstellar reddening since their
reddening line runs approximately along the sequences of dwarfs and
giants.  Taking into account that the scattering of color indices
$H$--$K_s$ due to intrinsic dispersion and observational errors is about
$\pm$\,0.05 mag (see Strai\v zys \& Lazauskait\.e 2009), the strip of
reddened RCGs (shown in Figure 1 in grey) is drawn with a 0.2 mag width
(in $H$--$K_s$).  It covers the main sequence in the G2--M2 range and
the giant branch in the G5--M5 range.  Luminosity class IV stars
(subgiants) of spectral types G5--K2\,IV should also fall within the
same strip.

Since G5--M dwarfs are absolutely faint, they are not seen at large
distances.  Contamination of the RCG strip from such stars can be
eliminated easily by setting an apparent magnitude limit in the $K_s$
vs.\,$H$--$K_s$ (magnitude-color) diagram.

However, contamination of the RCG strip by cooler giants is a serious
problem.  The giant sequence stars from K3\,III to M5\,III have more
negative $M_{K_s}$ magnitudes than RCGs, thus they are apparently
brighter and are visible at larger distances.  Although the space
density of RCGs outnumbers considerably that of K3--M5 III stars
(Perryman et al. 1995, 1997; Alves 2000), the surface density of the
latter is quite high.  Also, in the $J$--$H$ vs.\,$H$--$K_s$ diagram the
RCGs are mixed with giants of spectral types G5--K2 III which in the HR
diagram are on the ascending giant branch, i.e., burning hydrogen in a
shell around the helium core.  However, their luminosities in the $K_s$
passband do not differ considerably from those of RCGs, and their
presence in the sample does not distort the dependence of the extinction
on distance.  Hereafter, for simplicity, we will call all these stars as
RCGs.

Consequently, the main problem is the isolation of RCGs (together with
the intervening G5--K2 giants of the ascending branch) from cooler stars
of the red giant sequence, since {\it J, H, K} photometry alone cannot
help much.  We have to address the passbands of other photometric
systems.  Since we are usually dealing with faint stars, broad-band
systems are preferable.

\newpage

\sectionb{4}{SYNTHETIC COLORS IN THE u, g, r, i\, SYSTEM}

For the separation of RCGs from cooler giants we have applied the
$u$--$g$ vs.\,$r$--$i$ diagram where $u$ and $g$ are from the MegaCam
and $r$ and $i$ are from the IPHAS surveys.  Intrinsic color indices
$u$--$g$ and $r$--$i$ for the main-sequence stars, red giants and
F--G--K subdwarfs were calculated by the equation:
\begin{equation}
m_1-m_2 = - 2.5~{\rm {log}} {{\int
F(\lambda)\tau^x(\lambda)R_1(\lambda)d\lambda}
\over {\int F(\lambda)\tau^x(\lambda)R_2(\lambda)d\lambda}} + const,
\end{equation}
where $F(\lambda)$ are the flux distribution functions of stars in
energy units for the unit wavelength intervals, $\tau^x(\lambda)$ is the
transmittance function of $x$ units of interstellar dust (from
Strai\v{z}ys 1992), $R_1(\lambda)$ and $R_2(\lambda)$ are the
response-to-energy functions of passbands 1 and 2. The value of constant
makes all color indices of A0\,V star to be zeros.  Spectral energy
distributions for stars of luminosities V and III were taken from
Sviderskien\.e (1988) and for F--G--K subdwarfs from Sviderskien\.e
(1992).

Color indices can be used to calculate interstellar reddening-free
$Q$-parameters:
\begin{equation}
Q_{1234} = (m_1-m_2) - (E_{12}/E_{34})(m_3-m_4),
\end{equation}
here
\begin{equation}
E_{k,\ell} = (m_k-m_{\ell})_{\rm {reddened}} - (m_k-m_{\ell})_{\rm
{intrinsic}}.
\end{equation}

The response-to-energy functions, $R\,(\lambda)$, for the $r$ and $i$
filters in the IPHAS system were obtained by the equation:
\begin{equation}
R\,(\lambda) = \lambda\,f\,(\lambda)\,QE\,(\lambda),
\end{equation}
where $f\,(\lambda)$ are relative filter transmittances and {\it
QE}\,($\lambda$) is the quantum efficiency function of a CCD camera.
The filter transmittance and the quantum efficiency functions were taken
from the INT WFC and IPHAS Internet
sites.\footnote{~http://www.iphas.org and
http://www.ing.iac.es/Astronomy/instruments/wfc} The response-to-energy
functions of the MegaCam $u$ and $g$ magnitudes were obtained in a
similar way, but the filter transmittances and the MegaCam quantum
efficiency curve were taken from the MegaCam Internet
site.\footnote{~http://www2.cadc-ccda.hia-iha.nrc-cnrc.gc.ca/megapipe/docs/filters.html}

Since the observed MegaCam $u$ and $g$ magnitudes are in the AB
magnitude system\footnote {~For the definition of the system see Oke
(1965), AB means `ABsolute' flux.}, and our synthetic color indices
$u$--$g$ are normalized to zero for the A0\,V star (Vega), we have
corrected the synthetic indices by adding a constant of 0.438 mag.  The
$r$--$i$ synthetic color indices were used unchanged since the observed
IPHAS $r$ and $i$ magnitudes use the Vega-based zero-point.  The
$g$--$r$ synthetic color indices have mixed zero-point systems -- they
were corrected by subtracting 0.09 mag.  The values of the corrections
were taken from the MegaCam Internet site.  The accepted values of the
intrinsic color indices $u$--$g$, $g$--$r$ and $r$--$i$ are listed in
Table 1.

The slopes of reddening lines in the diagram $u$--$g$ vs.\,$r$--$i$
depend slightly on spectral type.  For early K-giants the ratio
$E_{u-g}/E_{r-i}$ is $\sim$\,1.50, for late K- and early M-giants it is
$\sim$\,1.45.  The ratios $E_{g-r}/E_{r-i}$ are 1.40 and 1.35,
respectively.


\begin{table}[!th]
\begin{center}
\vbox{\small\tabcolsep=4pt
\parbox[c]{110mm}{\baselineskip=9pt
{\smallbf Table 1.}{\small\ The synthetic intrinsic color indices
$u$--$g$, $g$--$r$ and $r$--$i$ accepted for luminosity V and III
stars and F--G subdwarfs. Zero-points of color indices $u$--$g$ are in
the MegaCam AB system, of $r$--$i$ in the IPHAS Vega system and of
$g$--$r$ in the mixed system (see the text). \lstrut}}
\begin{tabular}{lrrr|@{\huad}lccc}
\hline \hstrut
MK type  &  $u$--$g$ & $g$--$r$  & $r$--$i$ &
MK type  &  $u$--$g$ & $g$--$r$  & $r$--$i$ \\
\tablerule
 B0\,V  & --0.33 & --0.35 & --0.15 &  G5\,III & 1.37 & 0.73 & 0.45  \\ [-2pt]
 A0\,V  &  0.44  & --0.09 &  0.00  &  G8\,III & 1.53 & 0.79 & 0.47  \\ [-2pt]
 F0\,V  &  0.70  &   0.22 &  0.20  &  K0\,III & 1.69 & 0.85 & 0.50  \\ [-2pt]
 G0\,V  &  0.94  &   0.48 &  0.36  &  K1\,III & 1.82 & 0.92 & 0.54  \\ [-2pt]
 K0\,V  &  1.32  &   0.71 &  0.43  &  K2\,III & 1.99 & 1.00 & 0.57  \\ [-2pt]
 K2\,V  &  1.48  &   0.79 &  0.48  &  K3\,III & 2.28 & 1.07 & 0.58  \\ [-2pt]
 K3\,V  &  1.66  &   0.89 &  0.54  &  K4\,III & 2.42 & 1.20 & 0.67  \\ [-2pt]
 K5\,V  &  1.98  &   1.15 &  0.66  &  K5\,III & 2.61 & 1.31 & 0.80  \\ [-2pt]
 M0\,V  &  2.16  &   1.36 &  0.90  &  M0\,III & 2.73 & 1.33 & 0.90  \\ [-2pt]
 M2\,V  &  2.24  &   1.43 &  1.16  &  M2\,III & 2.76 & 1.39 & 1.06  \\ [-2pt]
 M4\,V  &  2.11  &   1.48 &  1.53  &  M3\,III & 2.70 & 1.42 & 1.29  \\ [-2pt]
 M5\,V  &  2.31  &   1.52 &  1.69  &  M4\,III & 2.55 & 1.43 & 1.62  \\ [-2pt]
 M7\,V  &  2.42  &   1.63 &  2.06  &  M5\,III & 2.35 & 1.51 & 1.96  \\ [-2pt]
 sd\,FG & 0.60 &   0.41 &  0.35  &  M6\,III & 2.05 & 1.65 & 2.32  \\
\tablerule
\end{tabular}
}
\end{center}
\end{table}


\begin{figure}[!th]
\vbox{
\centerline{\psfig{figure=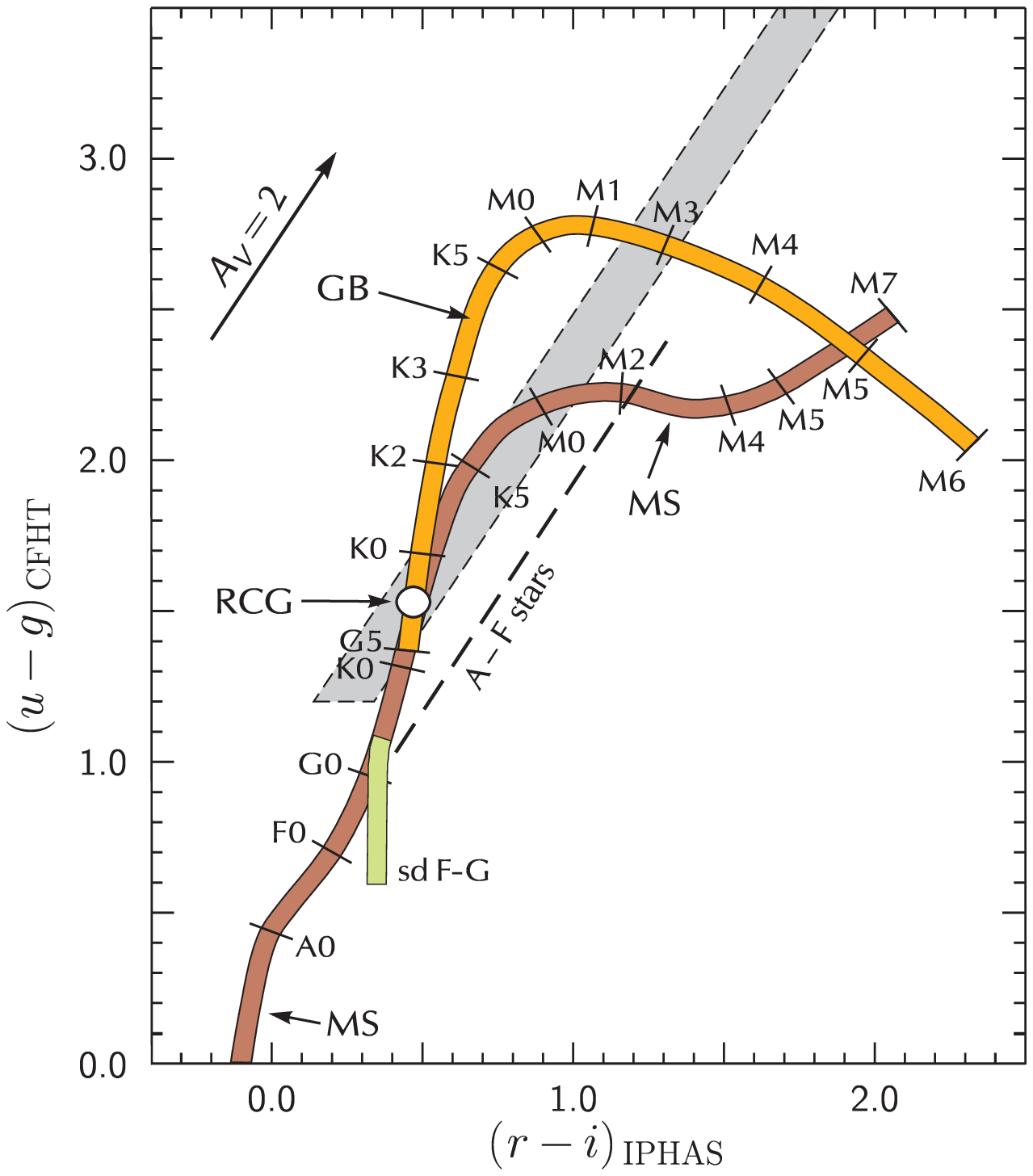,width=100mm,angle=0,clip=}}
\vspace{-.5mm}
\captionb{3}{The $u$--$g$ vs.\,$r$--$i$ diagram with the intrinsic main
sequence (MS, brown belt), the giant branch (GB, orange belt), the locus
of red clump giants (RCG, white circle) and the sequence of F--G
subdwarfs (sdF--G, green strip).  Two parallel broken lines mark the
boundaries of the grey strip in which RCGs with different interstellar
reddenings should be located.  The length of the reddening vector
corresponds to $A_V$ = 2 mag.}
}
\end{figure}

The schematic $u$--$g$ vs.\,$r$--$i$ diagram is shown in Figure 3. The
sequences for M-type stars of luminosity V and III classes are slightly
smoothed.  The exact position of RCGs in the diagram is unknown.  On the
analogy with the $J$--$H$ vs.\,$H$--$K_s$ diagram, their locus was taken
on the giant branch (GB), corresponding to spectral type G8\,III.  The
center line of the grey strip corresponds to
\begin{equation}
Q_{ugri} = (u-g) - 1.50\,(r-i) = 0.83,
\end{equation}
the boundary values are 0.68 and 0.98.  We accept that with
the increase of
interstellar reddening RCGs move upward and right along this strip.  The
strip extends over the main-sequence stars of spectral classes K0--M1
and crosses the giant branch at $\sim$\,M3\,III.  This means that this
diagram alone cannot isolate RCGs and intervening stars of spectral
classes G5--K1 of the giant sequence.

In an attempt to identify RCGs, we decided to combine the $J$--$H$
vs.\,$H$--$K_s$ (Figure 2) and $u$--$g$ vs.\,$r$--$i$ (Figure 3)
diagrams.  But first we have to identify the main features seen in the
observed diagrams of the NAP field.


\begin{figure}[!th]
\vbox{
\centerline{\psfig{figure=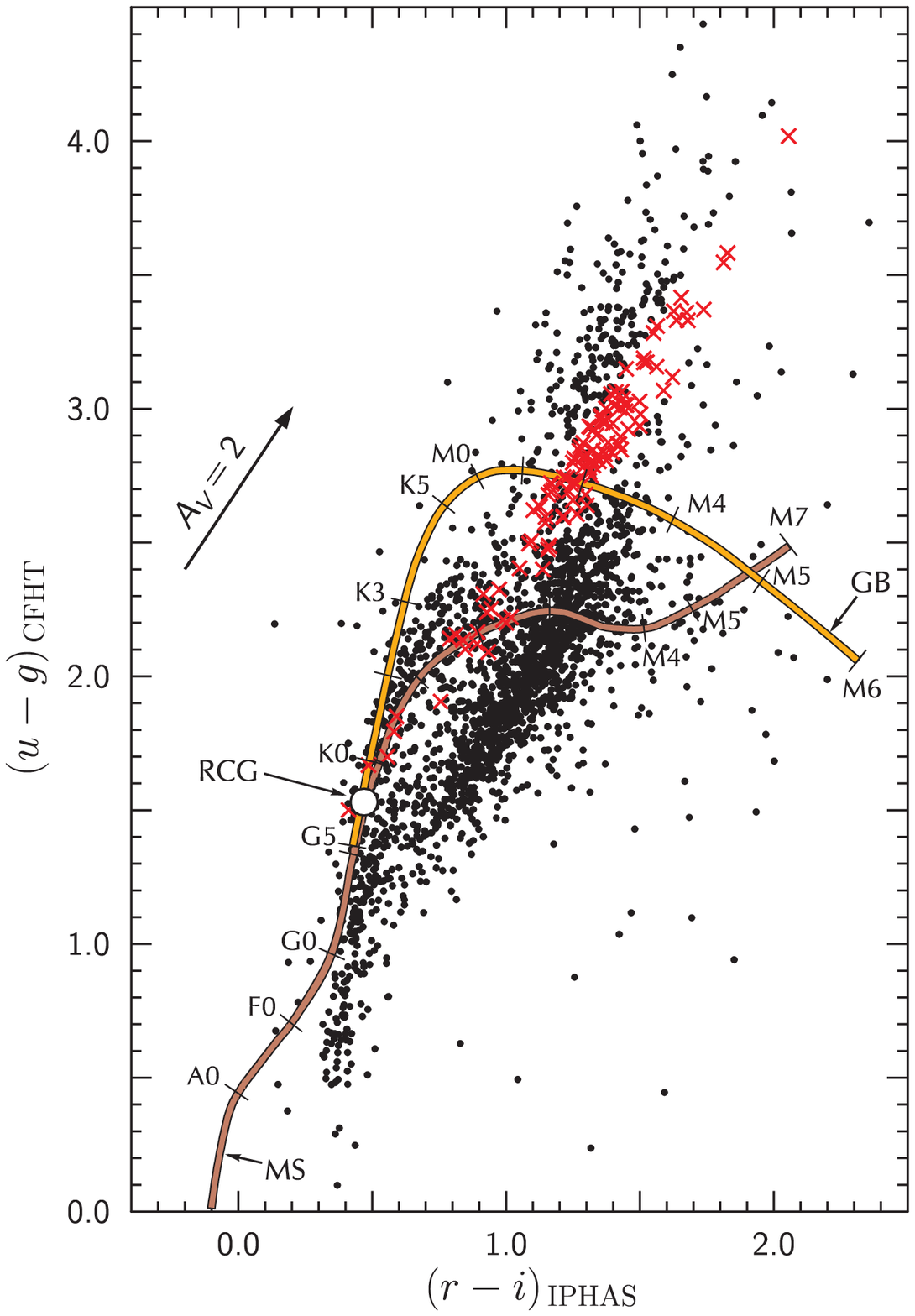,width=100mm,angle=0,clip=}}
\vspace{-.5mm}
\captionb{4}{Observed stars in the $u$--$g$ vs.\,$r$--$i$ diagram.  The
main sequence, giant branch and the locus of red clump giants are
overplotted from Figure 3. Red crosses denote the suspected RCGs (see
the text).}
}
\end{figure}

\sectionb{5}{THE OBSERVED u--g, r--i DIAGRAM}

In the NAP field shown in Figure 1 we identified about 2500 stars having
measurements in the MegaCam $u$ and $g$, the IPHAS $r$ and $i$ and the
2MASS {\it J, H, K}$_s$ magnitudes.  The magnitudes {\it u, g, r, i}
were taken up to an accuracy of 0.1 mag, while for the 2MASS magnitudes
the accuracy limit was 0.05 mag.  Figure 4 shows the $u$--$g$
vs.\,$r$--$i$ diagram for the selected sample together with the
sequences of luminosity V and III stars and the interstellar reddening
vector.

The distribution of stars in this diagram is the combined result of real
distribution of stars on the intrinsic sequences of different luminosity
classes, different shifts of stars along their reddening lines, and the
limiting magnitudes both at the bright and faint ends of different
passbands.  Limiting magnitudes depend not only on the exposure lengths
but also on the temperature, luminosity and interstellar reddening.  At
the faint end the limiting magnitude is defined mainly by the
signal-to-noise ({\it S/N}) ratio and the accepted accuracy limit, at
the bright end it is defined by the saturation of images in CCD
detectors.  Among the four passbands used (Figure 4) the worst {\it S/N}
ratio is for the $u$ magnitude since in the ultraviolet the CCD+filter
sensitivity is the lowest and in this spectral region cool and heavily
reddened stars (which are the objects of our interest) are quite faint.


\begin{figure}[!t]
\centerline{\hbox{
\psfig{figure=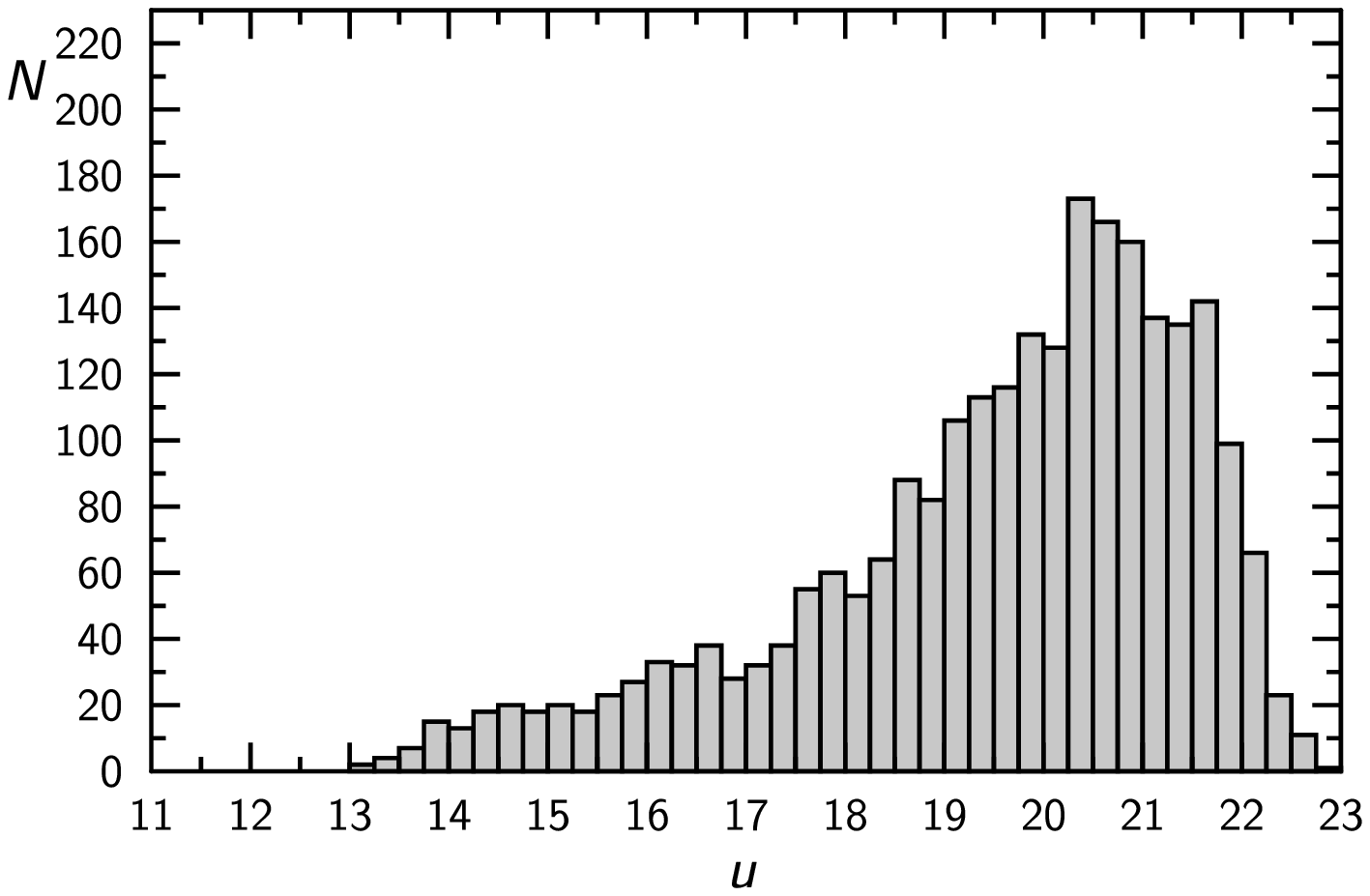,width=61mm,angle=0,clip=}
\hskip1mm
\psfig{figure=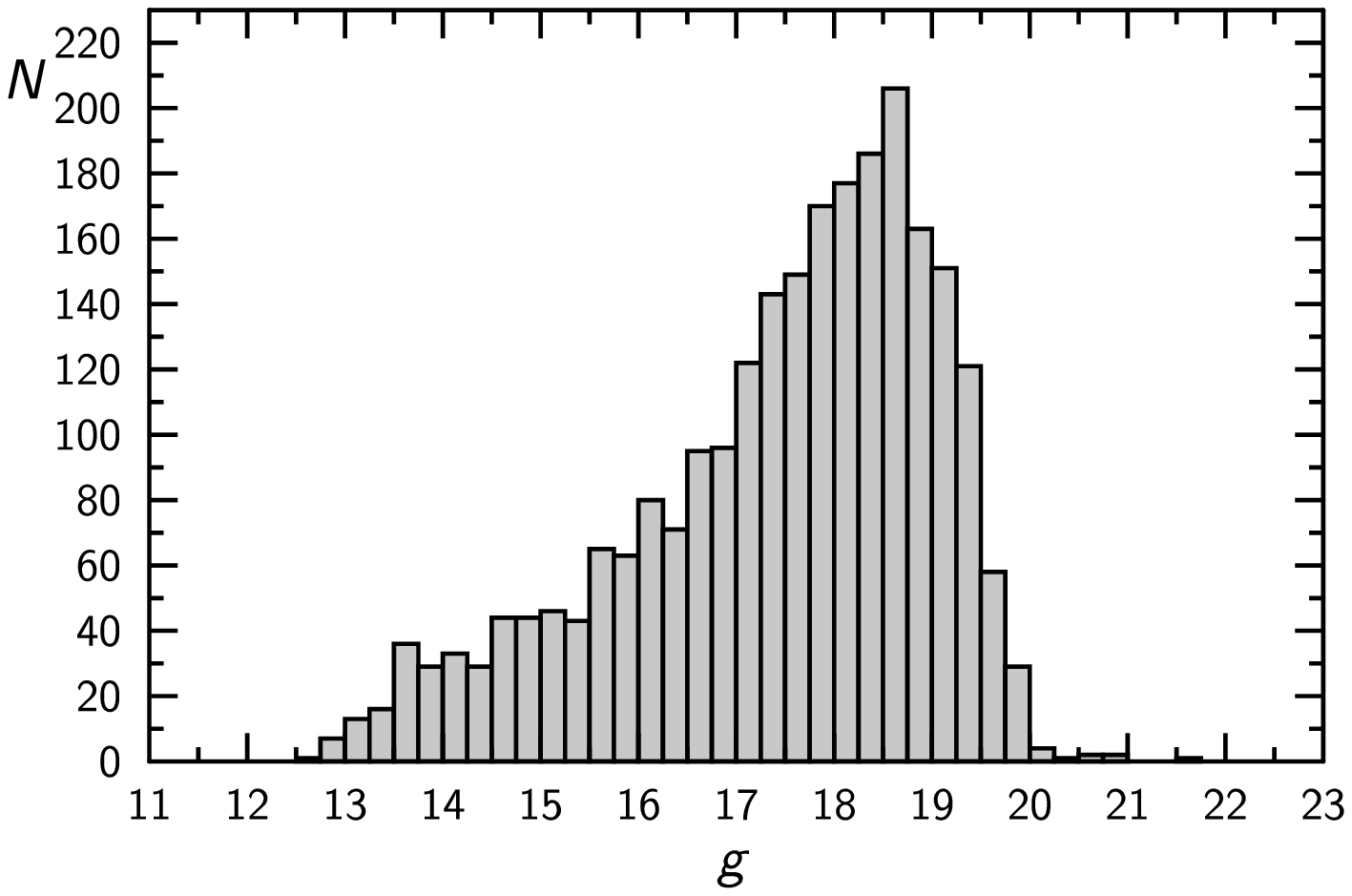,width=61mm,angle=0,clip=}}}
\vskip3mm
\hbox{
\centerline{\psfig{figure=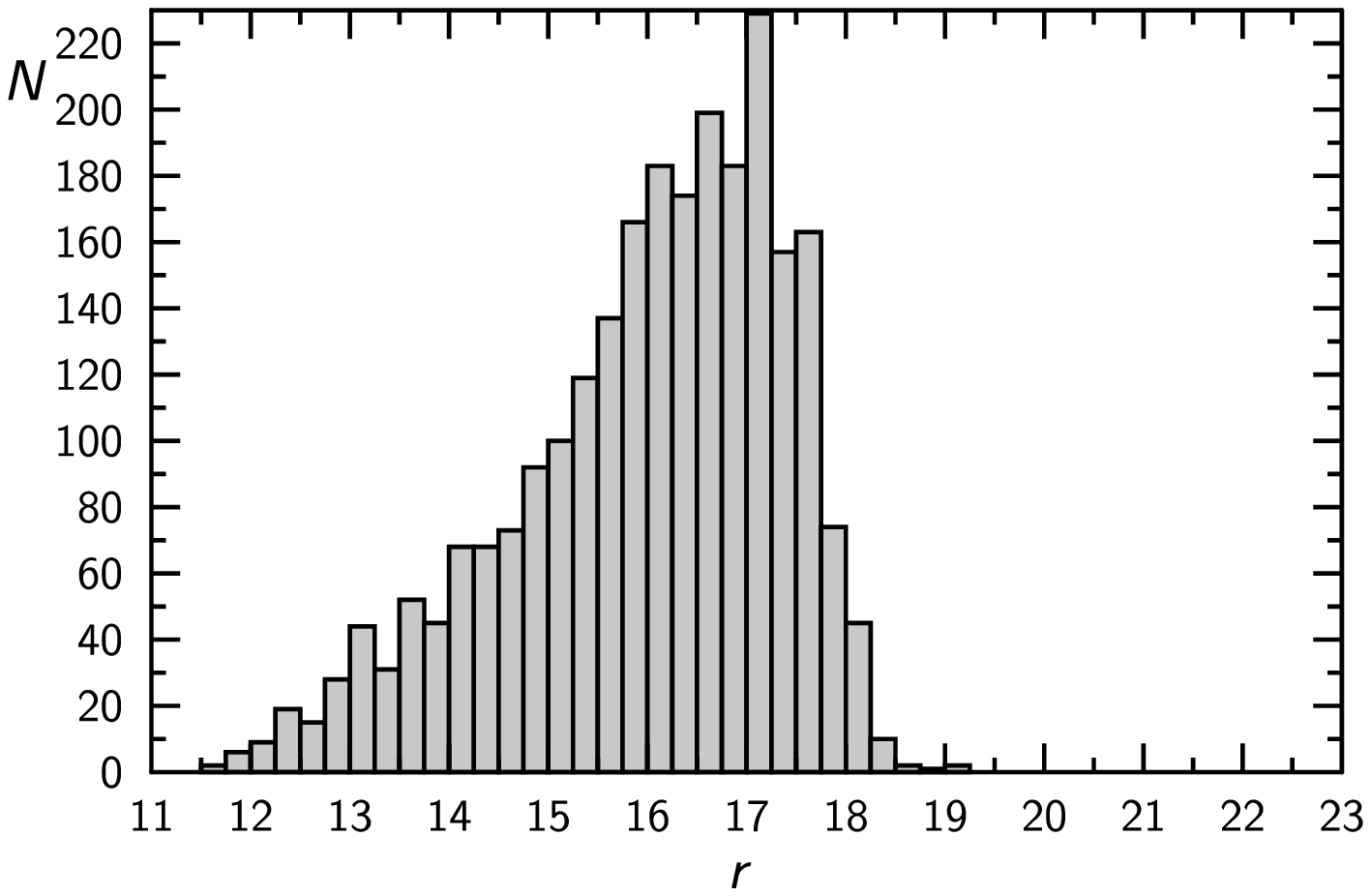,width=61truemm,angle=0,clip=}
\hskip1mm
\psfig{figure=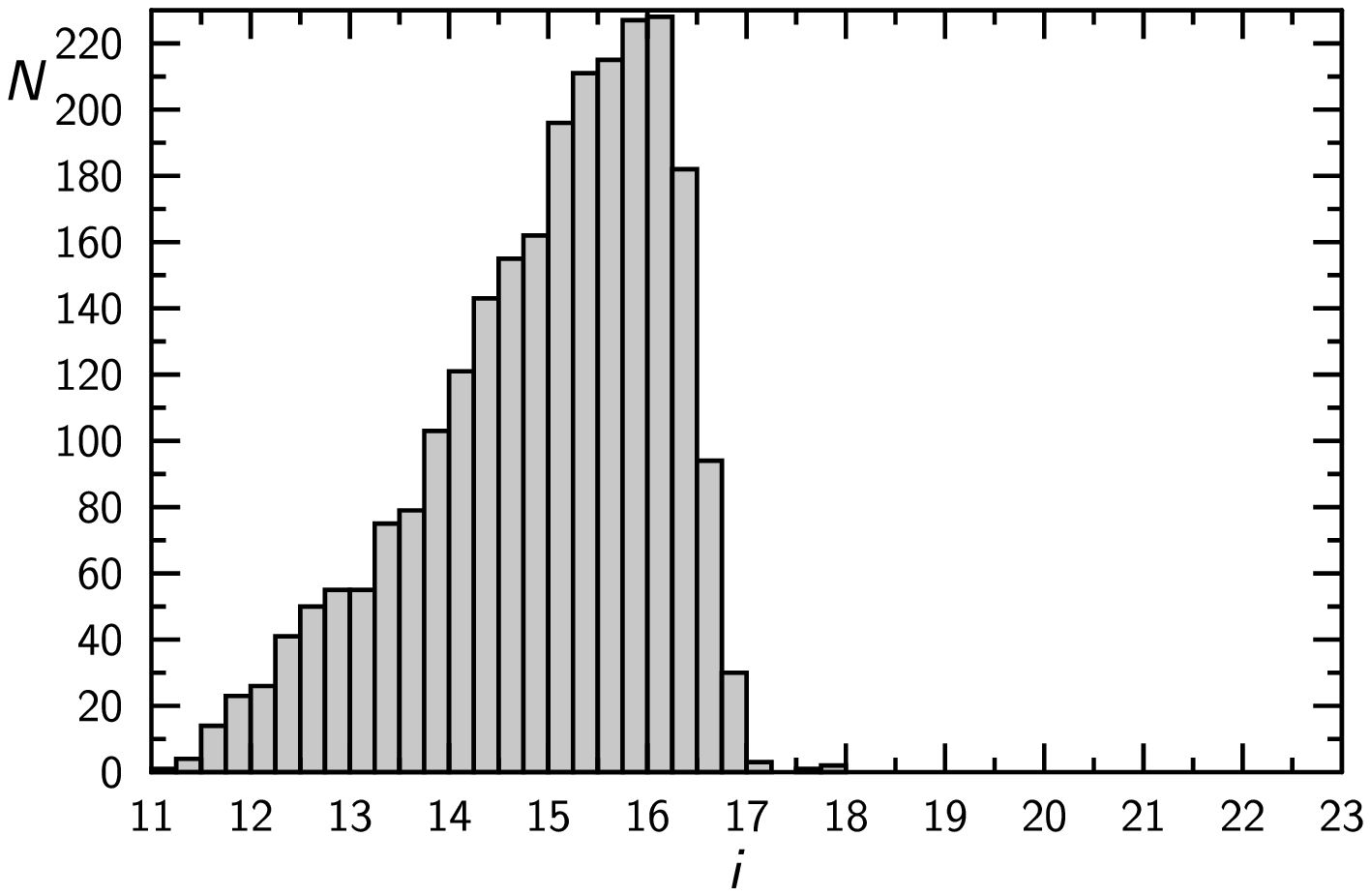,width=61truemm,angle=0,clip=}}}
\vskip1mm
\captionb{5}{Distributions by magnitudes $u$, $g$, $r$ and $i$ of the
2500 stars in the selected sample.}
\end{figure}

Figure 5 shows the distribution of stars in our sample with respect to
magnitude in the four passbands.  The faintest limiting magnitudes are
close to $u$ = 22, $g$ = 20, $r$ = 18 and $i$ = 16.5.  The brightest end
for $u$ and $g$ magnitudes is close to 13 and for $r$ and $i$ close to
12.  This means that in our sample all stars brighter than $r$ = 12 are
missing, and this magnitude cut eliminates from the sample some
types of stars at small distances from the Sun, such as B, A and F stars
up to a few hundreds of parsecs.

At first glance, we can spot in the $u$--$g$ vs.\,$r$--$i$ diagram some
outstanding features.  First, we see a well populated intrinsic sequence
of G, K and early M dwarfs.  The coolest dwarfs are absolutely faint,
especially in the ultraviolet, and they are rare in the selected sample.
The stars of spectral classes B, A and F are absent on the intrinsic
main sequence since they are too bright in $u$ and $g$.  The stars of
latest B subclasses and of all A subclasses form a rich-populated
sequence with different reddenings which extends from $r$--$i$ = 0.4 to
1.5 or even farther to the red.  Due to their high luminosity and
relatively high space density these stars are seen at large distances.
Except for A-type stars, whose intrinsic sequence runs almost along the
reddening line, the sequence of reddened stars should also contain
F-type stars, but due to their lower luminosity they are expected to be
numerous only at small distances and reddenings.

The second rich sequence, running more or less in parallel  with the
sequence of A stars and lying above it, corresponds to K--M giants,
including RCGs -- it extends from $r$--$i$ = 0.5 to $\sim$\,2.0.  This
sequence is broader since the intrinsic line of red giants runs at a
considerable angle with respect to the direction of reddening lines.
The sequence includes a sample of stars shown as red crosses.  At
present it is sufficient to say that these stars are supposed RCGs.  For
the procedure of their identification see Section 7.

The next feature seen in the $u$--$g$ vs.\,$r$--$i$ diagram is the
strip of possible F--G metal-deficient dwarfs (subdwarfs) at $r$--$i$
= 0.35.  Most of them should be located in front of the NAP complex and
therefore have low reddenings. In the same region of the diagram and
farther to the right we could find also reddened B-type stars of early
subclasses, if such were present in the sample. However, it is easy to
separate reddened B stars from unreddened F--G subdwarfs in the
$J$--$H$ vs.\,$H$--$K_s$ diagram. None star located in this strip has
been found to be a reddened B star.


\begin{figure}[!th]
\vbox{
\centerline{\psfig{figure=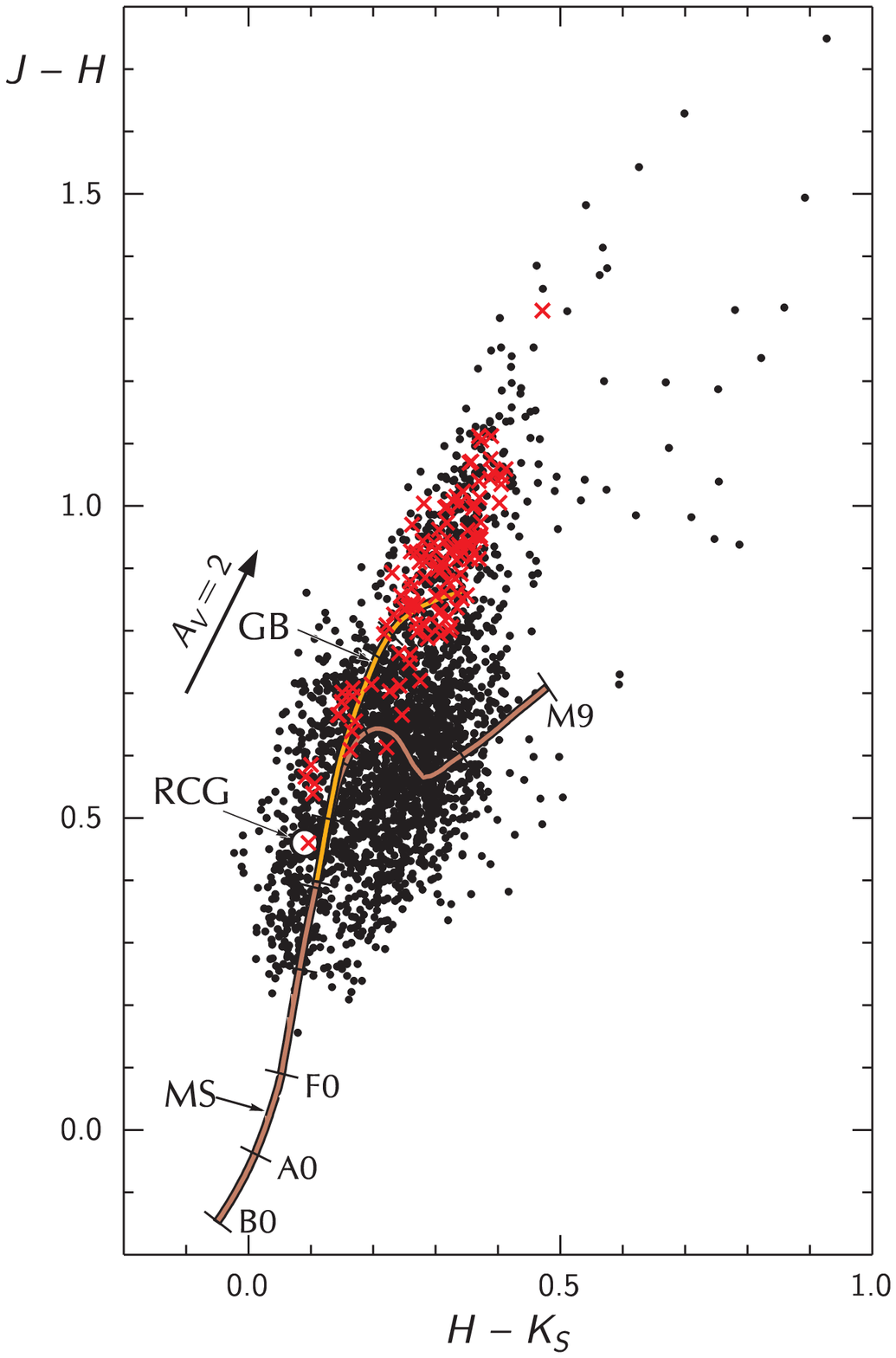,width=120mm,angle=0,clip=}}
\vspace{-.5mm}
\captionb{6}{Observed stars in the $J$--$H$ vs.\,$H$--$K_s$ diagram.
The main sequence, giant branch and the locus of red clump giants
are overplotted from Figure 2. Red crosses denote the
suspected RCGs.}
}
\vskip-5mm
\end{figure}

\sectionb{6}{THE OBSERVED J--H vs.\,H--K$_s$ DIAGRAM}

The $J$--$H$ vs.\,$H$--$K_s$ diagram for the same list of stars is
plotted in Figure 6. The stars lower than $J$--$H$\,$<$\,0.2 are absent
due to the saturation effect in the $u$ and partly $g$ images of the
MegaCam system.  At larger values of $J$--$H$ crowding of stars and
ovelapping of the reddened sequences is much larger than it was observed
in the $u$--$g$ vs.\,$r$--$i$ diagram.  Comparing Figures 6 and 2, we
can identify some of the features seen in the $u$--$g$ vs.\,$r$--$i$
diagram.  The sequence of unreddened G dwarfs is seen only from $J$--$H$
= 0.25 to 0.4 (G0--K0 V).  K- and M-dwarfs are completely overlapped
with giants.  The largest concentration of reddened A stars is moved
onto the M-dwarf sequence and to both sides of it.  The sequence of
reddened K--M giants is seen more distinctly, being narrower than in the
$u$--$g$ vs.\,$r$--$i$ diagram.  However, as it is evident from Figure
2, K and M giants of different reddenings, including RCGs (shown as red
crosses), overlap each other.

In Figure 6 there are seen two groups of 44 dots deviating considerably
to the right of the described sequences, and having $Q_{JHK_s}$ between
--0.2 and --0.8.  This is a typical domain of pre-main-sequence stars
(PMSs or YSOs) where Herbig Ae/Be and T Tauri stars are located (see
Figures 5 and 6 in Corbally et al. 2009).  T Tauri stars can be present
only in the upper group with $J$--$H$\,$>$\,0.9 and
$H$--$K_s$\,$>$\,0.6.  Two of them are known T Tauri stars, V\,1539 Cyg
and V\,521 Cyg.  The lower group of points probably contains massive
YSOs of Herbig Ae/Be type.  Some heavily reddened Herbig stars can be
also present in the upper group.  However, we do not exclude the
possibility that some of these objects can be AGB stars, point-like
galaxies or quasars.

\sectionb{7}{ISOLATION OF RED CLUMP GIANTS}

For the isolation of RCGs (together with G5--K1 giants on the
ascending branch), we adopted several conditions ensuring the exclusion
of overlapping stars from other sequences.

1. In the $J$--$H$ vs.\,$H$--$K_s$ diagram the requirement was set that
the star must lie inside the grey strip shown in Figure 2, outlined by
the reddening lines with $Q_{JHK_s}$ between 0.18 and 0.38.

2. In the $u$--$g$ vs.\,$r$--$i$ diagram, the requirement was set that
the star must lie inside the grey strip shown in Figure 3, outlined by
the reddening lines with $Q_{ugri}$ between 0.68 and 0.98. This
requirement separates RCG stars from giants cooler than K1.
Since the strip crosses the red giant sequence at class M3, it is
impossible to separate reddened RCGs from stars of types M2--M3\,III.

3. Again in the $J$--$H$ vs.\,$H$--$K_s$ diagram, in an attempt to
exclude the main sequence stars earlier than $\sim$\,K3\,V and giants
hotter than RCGs, we accepted the following two conditions:
$J$--$H$\,$>$\,0.46 and $H$--$K_s$\,$>$\,0.09.  These color cuts
correspond to the intrinsic colors of RCGs, and they preclude the
appearance between RCGs of stars with negative color excesses.

4. Even with these four conditions met, the selected RCG sample includes
an admixture of the main-sequence stars -- unreddened dwarfs of types
K3--M1 V and reddened dwarfs of types K0--M1 V. To exclude most of them,
we added one more condition, $K$\,$<$\,12 + 2.0\,($H$--$K_s$), specific
for the NAP region.  This restriction cuts the red dwarfs of types G5\,V
and cooler, located at and beyond the distance of the NAP complex.

Applying these conditions, we have selected a sample of 104 stars,
listed in Table 2, which are probable RCGs with an admixture of normal
giant branch stars of spectral classes G5--K1 III and $\sim$\,M3\,III.
In Figures 4 and 6 these stars are plotted as red crosses.  All these
stars lie within the grey strips shown in Figures 2 and 3. However,
within these strips we find also many unreddened and reddened
main-sequence stars (and subgiants), mostly of spectral classes G8--K,
which do not satisfy conditions (3) and (4).

\newpage

{\small
\tabcolsep=3pt
\begin{longtable}{rccrccccccc}
\multicolumn{11}{l}{\parbox{115mm}{\baselineskip=9pt
{\bf\ \ Table 2.}{\ Stars in the NAP area with photometric properties of
red clump giants in the 2MASS, IPHAS and MegaCam systems. 17 stars with
the numbers flagged with asterisks probably have luminosities lower than
those of RCGs.  For these stars the extinction and distance values are
not given.\lstrut}}}\\
\firsthline
\noalign{\vskip1mm}
No.& RA\,(2000) & DEC\,(2000) & $K_s$~~ & $J$--$H$ & $H$--$K_s$ &$u$--$g$ & $g$--$r$ & $r$--$i$ & $A_V$ & $d$\,(kpc) \\
\noalign{\vskip1mm}
\lasthline
\endfirsthead

\noalign{\vskip1mm}
\multicolumn{11}{l}{{\bf\ \ Table 2.}{\ Continued\lstrut}}\\
\hline
\noalign{\vskip1mm}
No.& RA\,(2000) & DEC\,(2000) & $K_s$~~ & $J$--$H$ & $H$--$K_s$ &$u$--$g$ & $g$--$r$ & $r$--$i$ & $A_V$ & $d$\,(kpc) \\
\noalign{\vskip1mm}
\lasthline
\endhead

\endfoot
  1\rlap{*}& 313.566	&43.729	&11.022	&0.555	&0.106	&1.793	&1.156	&0.582	& &  \\
  2\rlap{*}& 313.583	&43.438	&11.664	&0.639	&0.166	&2.101	&1.306	&0.848	&     &    \\
  3 & 313.635	&43.309	&9.963	&1.069	&0.354	&3.582	&1.989	&1.827	&4.38	&1.61 \\
  4\rlap{*}& 313.661	&43.564	&11.674	&0.539	&0.103	&1.700	&1.000	&0.557	&     &    \\
  5 & 313.665	&43.504	&9.362	&1.059	&0.412	&3.370	&2.308	&1.738	&5.34	&1.15 \\
  6 & 313.687	&43.278	&11.706	&0.927	&0.324	&3.012	&1.969	&1.405	&3.88	&3.69 \\
  7 & 313.687	&43.330	&9.015	&0.905	&0.312	&2.951	&1.877	&1.373	&3.69	&1.08 \\
  8\rlap{*}& 313.693	&44.233	&11.682	&0.609	&0.163	&1.906	&1.038	&0.756	&      &    \\
  9 & 313.726	&43.384	&10.281	&1.044	&0.387	&3.415	&2.206	&1.655	&4.93	&1.80 \\
 10\rlap{*}& 313.734	&43.339	&12.148	&0.626	&0.149	&2.009	&1.138	&0.732	&    &      \\
 11\rlap{*}& 313.791	&43.379	&11.278	&0.695	&0.167	&2.165	&1.263	&0.894	&     &    \\
 12 & 313.792	&43.274	&11.229	&0.807	&0.305	&2.949	&1.924	&1.361	&3.57	&3.01 \\
 13\rlap{*}& 313.830	&44.241	&10.694	&0.665	&0.143	&2.155	&1.226	&0.813	&     &     \\
 14 & 313.844	&43.522	&7.837	&1.105	&0.373	&3.547	&2.520	&1.813	&4.70	&0.59  \\
 15 & 313.914	&44.048	&7.537	&1.313	&0.471	&4.019	&2.711	&2.056	&6.33	&0.47  \\
 16\rlap{*}& 313.934	&44.240	&12.266	&0.632	&0.160	&2.083	&1.129	&0.769	&    &      \\
 17 & 313.971	&44.212	&8.173	&1.055	&0.392	&3.283	&2.178	&1.551	&5.01	&0.68  \\
 18 & 313.981	&44.147	&8.304	&1.035	&0.405	&3.330	&2.281	&1.679	&5.23	&0.71  \\
 19\rlap{*}& 314.007	&44.021	&9.891	&0.665	&0.145	&2.141	&1.298	&0.790	&     &      \\
 20 & 314.032	&44.233	&10.897	&0.891	&0.340	&2.639	&1.828	&1.302	&4.15	&2.50 \\
 21 & 314.034	&44.244	&10.524	&0.953	&0.370	&3.014	&2.064	&1.444	&4.65	&2.05 \\
 22 & 314.059	&43.326	&9.087	&1.040	&0.369	&3.360	&2.215	&1.674	&4.63	&1.06  \\
 23 & 314.067	&44.149	&11.089	&0.941	&0.297	&2.902	&1.923	&1.338	&3.44	&2.85 \\
 24 & 314.080	&44.076	&11.708	&0.798	&0.270	&2.486	&1.590	&1.162	&2.99	&3.88 \\
 25 & 314.090	&44.243	&9.794	&0.938	&0.337	&2.867	&1.995	&1.423	&4.10	&1.51  \\
 26 & 314.098	&44.258	&9.835	&1.075	&0.388	&3.330	&2.324	&1.637	&4.95	&1.47  \\
 27 & 314.100	&44.056	&10.279	&0.832	&0.308	&2.692	&1.778	&1.234	&3.62	&1.94 \\
 28 & 314.133	&44.264	&9.560	&0.909	&0.309	&2.814	&1.872	&1.255	&3.63	&1.39  \\
 29 & 314.137	&43.981	&11.133	&0.920	&0.340	&3.003	&1.968	&1.425	&4.15	&2.79 \\
 30 & 314.176	&44.159	&11.102	&0.914	&0.333	&2.720	&1.830	&1.263	&4.03	&2.77 \\
 31\rlap{*}& 314.215	&43.714	&12.048	&0.619	&0.165	&1.901	&1.043	&0.713	&      &    \\
 32\rlap{*}& 314.216	&43.387	&11.538	&0.714	&0.197	&2.121	&1.076	&0.892	&     &     \\
 33 & 314.221	&44.273	&12.262	&0.809	&0.282	&2.767	&1.876	&1.308	&3.19	&4.96 \\
 34 & 314.223	&44.229	&11.035	&0.899	&0.302	&2.782	&1.726	&1.288	&3.52	&2.76 \\
 35 & 314.225	&44.136	&11.823	&0.934	&0.339	&2.806	&1.913	&1.375	&4.13	&3.84 \\
 36\rlap{*}& 314.243	&43.300	&11.285	&0.708	&0.167	&2.220	&1.324	&1.019	&    &      \\
 37 & 314.283	&43.699	&11.088	&0.973	&0.318	&2.744	&2.044	&1.215	&3.79	&2.79 \\
 38 & 314.301	&44.252	&10.898	&0.935	&0.330	&2.830	&1.864	&1.369	&3.98	&2.53 \\
 39 & 314.332	&44.210	&10.011	&1.013	&0.370	&3.068	&2.178	&1.589	&4.65	&1.62 \\
 40 & 314.338	&44.068	&10.755	&0.915	&0.356	&3.062	&1.944	&1.410	&4.42	&2.31 \\
 41 & 314.388	&43.577	&10.063	&0.974	&0.372	&3.364	&2.126	&1.627	&4.68	&1.65 \\
 42 & 314.402	&44.067	&10.512	&1.004	&0.338	&3.156	&2.030	&1.563	&4.12	&2.10 \\
 43 & 314.428	&43.254	&10.320	&0.911	&0.284	&2.817	&1.827	&1.286	&3.22	&2.02 \\
 44\rlap{*}& 314.443	&43.305	&9.409	&0.655	&0.171	&2.136	&1.451	&0.839	&      &     \\
 45 & 314.458	&44.160	&11.139	&1.023	&0.344	&3.118	&2.153	&1.622	&4.22	&2.79 \\
 46 & 314.471	&43.442	&10.956	&0.929	&0.300	&2.750	&1.791	&1.318	&3.49	&2.67 \\
 47 & 314.476	&44.212	&11.488	&0.997	&0.315	&3.173	&2.003	&1.516	&3.73	&3.36 \\
 48 & 314.491	&43.502	&10.264	&0.879	&0.259	&2.859	&1.785	&1.276	&2.81	&2.01 \\
 49 & 314.503	&44.216	&9.460	&0.924	&0.283	&2.841	&1.736	&1.364	&3.20	&1.36  \\
 50 & 314.522	&43.423	&11.188	&0.809	&0.221	&2.599	&1.595	&1.213	&2.17	&3.20 \\
 51 & 314.529	&43.419	&12.060	&0.811	&0.268	&2.595	&1.671	&1.148	&2.96	&4.57 \\
 52 & 314.531	&43.581	&12.282	&0.884	&0.331	&2.973	&1.901	&1.356	&4.00	&4.78 \\
 53 & 314.534	&43.520	&11.579	&0.712	&0.242	&2.307	&1.302	&0.916	&2.52	&3.75 \\
 54 & 314.541	&43.291	&11.578	&0.862	&0.313	&2.799	&1.806	&1.250	&3.70	&3.51 \\
 55 & 314.545	&43.886	&10.869	&1.047	&0.406	&2.923	&1.194	&1.457	&5.25	&2.33 \\
 56 & 314.547	&43.259	&12.013	&0.788	&0.287	&2.399	&1.612	&1.137	&3.27	&4.40 \\
 57 & 314.553	&44.138	&10.289	&0.962	&0.305	&3.038	&1.931	&1.438	&3.57	&1.95 \\
 58 & 314.555	&43.436	&11.254	&0.840	&0.263	&2.474	&1.536	&1.157	&2.87	&3.17 \\
 59 & 314.574	&43.407	&11.480	&0.843	&0.264	&2.688	&1.644	&1.169	&2.89	&3.51 \\
 60 & 314.596	&44.273	&10.084	&0.943	&0.313	&3.001	&1.975	&1.373	&3.70	&1.76 \\
 61\rlap{*}& 314.598	&43.555	&12.243	&0.515	&0.146	&1.608	&0.830	&0.533	&    &      \\
 62 & 314.605	&44.136	&10.686	&0.880	&0.319	&3.063	&1.888	&1.432	&3.80	&2.32 \\
 63 & 314.615	&44.162	&11.264	&0.839	&0.256	&2.722	&1.692	&1.256	&2.76	&3.20 \\
 64 & 314.621	&43.451	&12.517	&0.864	&0.299	&2.673	&1.746	&1.202	&3.47	&5.49 \\
 65 & 314.637	&43.438	&10.614	&0.868	&0.261	&2.719	&1.667	&1.171	&2.84	&2.36 \\
 66 & 314.652	&44.227	&9.541	&0.938	&0.353	&3.013	&1.863	&1.450	&4.37	&1.32  \\
 67\rlap{*}& 314.665	&44.018	&11.857	&0.460	&0.096	&1.500	&0.700	&0.412	&    &      \\
 68 & 314.671	&43.532	&10.389	&0.860	&0.335	&2.606	&1.719	&1.265	&4.07	&1.99 \\
 69 & 314.675	&43.548	&12.127	&0.943	&0.352	&2.805	&1.916	&1.324	&4.35	&4.37 \\
 70 & 314.681	&43.650	&10.498	&0.748	&0.258	&2.324	&1.408	&0.975	&2.79	&2.25 \\
 71\rlap{*}& 314.687	&44.134	&12.135	&0.499	&0.135	&1.672	&0.725	&0.549	&     &     \\
 72 & 314.695	&43.258	&11.629	&0.999	&0.318	&2.966	&1.950	&1.357	&3.79	&3.58 \\
 73 & 314.705	&43.597	&11.374	&0.997	&0.360	&3.055	&2.021	&1.392	&4.48	&3.06 \\
 74 & 314.732	&43.487	&10.966	&0.846	&0.262	&2.697	&1.675	&1.176	&2.86	&2.78 \\
 75 & 314.733	&43.528	&11.473	&0.925	&0.362	&2.868	&1.848	&1.418	&4.51	&3.20 \\
 76 & 314.753	&44.234	&9.533	&1.112	&0.389	&3.173	&1.997	&1.526	&4.96	&1.27  \\
 77 & 314.763	&43.505	&12.042	&0.762	&0.259	&2.505	&1.618	&1.095	&2.81	&4.57 \\
 78\rlap{*}& 314.773	&43.665	&12.237	&0.621	&0.155	&1.853	&1.071	&0.647	&    &      \\
 79 & 314.773	&43.323	&12.279	&0.829	&0.305	&2.932	&1.893	&1.315	&3.57	&4.89 \\
 80 & 314.777	&43.952	&10.712	&1.002	&0.362	&2.980	&2.074	&1.503	&4.51	&2.25 \\
 81 & 314.798	&43.322	&11.187	&0.952	&0.371	&3.309	&2.134	&1.565	&4.67	&2.78 \\
 82 & 314.821	&43.483	&11.985	&0.918	&0.290	&2.843	&1.805	&1.274	&3.32	&4.33 \\
 83 & 314.825	&44.134	&12.226	&0.843	&0.238	&2.414	&1.586	&1.030	&2.46	&5.08 \\
 84 & 314.842	&43.338	&11.153	&0.961	&0.351	&3.149	&2.031	&1.448	&4.33	&2.79 \\
 85 & 314.847	&44.225	&12.183	&1.005	&0.402	&2.779	&2.030	&1.310	&5.18	&4.28 \\
 86 & 314.858	&43.456	&11.208	&0.910	&0.274	&2.821	&1.826	&1.300	&3.05	&3.07 \\
 87 & 314.858	&44.196	&10.491	&0.947	&0.372	&2.716	&1.983	&1.245	&4.68	&2.02 \\
 88 & 314.860	&43.346	&10.522	&0.952	&0.363	&3.188	&2.117	&1.515	&4.53	&2.06 \\
 89 & 314.870	&44.164	&10.877	&1.008	&0.328	&2.847	&1.794	&1.430	&3.95	&2.51 \\
 90 & 314.877	&43.382	&11.558	&0.885	&0.283	&2.872	&1.804	&1.288	&3.20	&3.58 \\
 91 & 314.890	&44.130	&12.260	&1.015	&0.333	&3.027	&1.581	&1.499	&4.03	&4.72 \\
 92 & 314.900	&43.502	&9.450	&0.856	&0.244	&2.634	&1.504	&1.128	&2.56	&1.40  \\
 93 & 314.903	&44.122	&10.433	&0.826	&0.233	&2.488	&1.452	&1.159	&2.37	&2.23 \\
 94 & 314.905	&43.411	&10.127	&0.856	&0.300	&2.743	&1.662	&1.228	&3.49	&1.82 \\
 95 & 314.906	&43.800	&11.692	&0.811	&0.279	&2.496	&1.953	&1.086	&3.14	&3.82 \\
 96 & 314.909	&43.549	&11.567	&0.841	&0.271	&2.728	&1.717	&1.252	&3.00	&3.63 \\
 97 & 314.909	&43.418	&11.095	&0.896	&0.309	&2.934	&1.838	&1.310	&3.63	&2.82 \\
 98 & 314.910	&43.933	&11.222	&1.071	&0.357	&2.803	&2.373	&1.347	&4.43	&2.86 \\
 99 & 314.911	&44.258	&11.909	&0.835	&0.250	&2.574	&1.581	&1.147	&2.66	&4.34 \\
100 & 314.912	&44.183	&10.717	&1.112	&0.369	&2.836	&2.096	&1.396	&4.63	&2.24 \\
101 & 314.916	&43.378	&10.050	&0.795	&0.217	&2.678	&1.678	&1.158	&2.11	&1.90 \\
102 & 314.923	&43.483	&10.960	&0.703	&0.226	&2.091	&1.303	&0.935	&2.26	&2.86 \\
103 & 314.924	&43.482	&12.371	&0.801	&0.239	&2.539	&1.597	&1.128	&2.47	&5.42 \\
104 & 314.926	&43.308	&8.900	&0.764	&0.241	&2.593	&1.643	&1.213	&2.51	&1.09  \\
\tablerule
\end{longtable}
}

\sectionb{8}{THE OBSERVED g--r, r--i DIAGRAM}

Figure 7 shows the $g$--$r$ vs.\,$r$--$i$ diagram for the same sample of
stars.  The sequences of luminosity V and III stars shown in the diagram
were taken from Table 1, they correspond to the $g$--$r$ index combined
from two photometric systems, MegaCam and IPHAS, with different zero
points.  In this diagram we see the same features as in the $u$--$g$
vs.\,$r$--$i$ diagram:  the unreddened main sequence of G--K--M dwarfs
and the rich sequence of reddened A and F stars on the extension of the
intrinsic main sequence of these spectral classes.  The upper part of
this sequence is overlapped by the sequence of reddened RCGs (red
crosses).  Above it a broad band of reddened K2--K5 giants is located.


\begin{figure}[!th]
\vbox{
\centerline{\psfig{figure=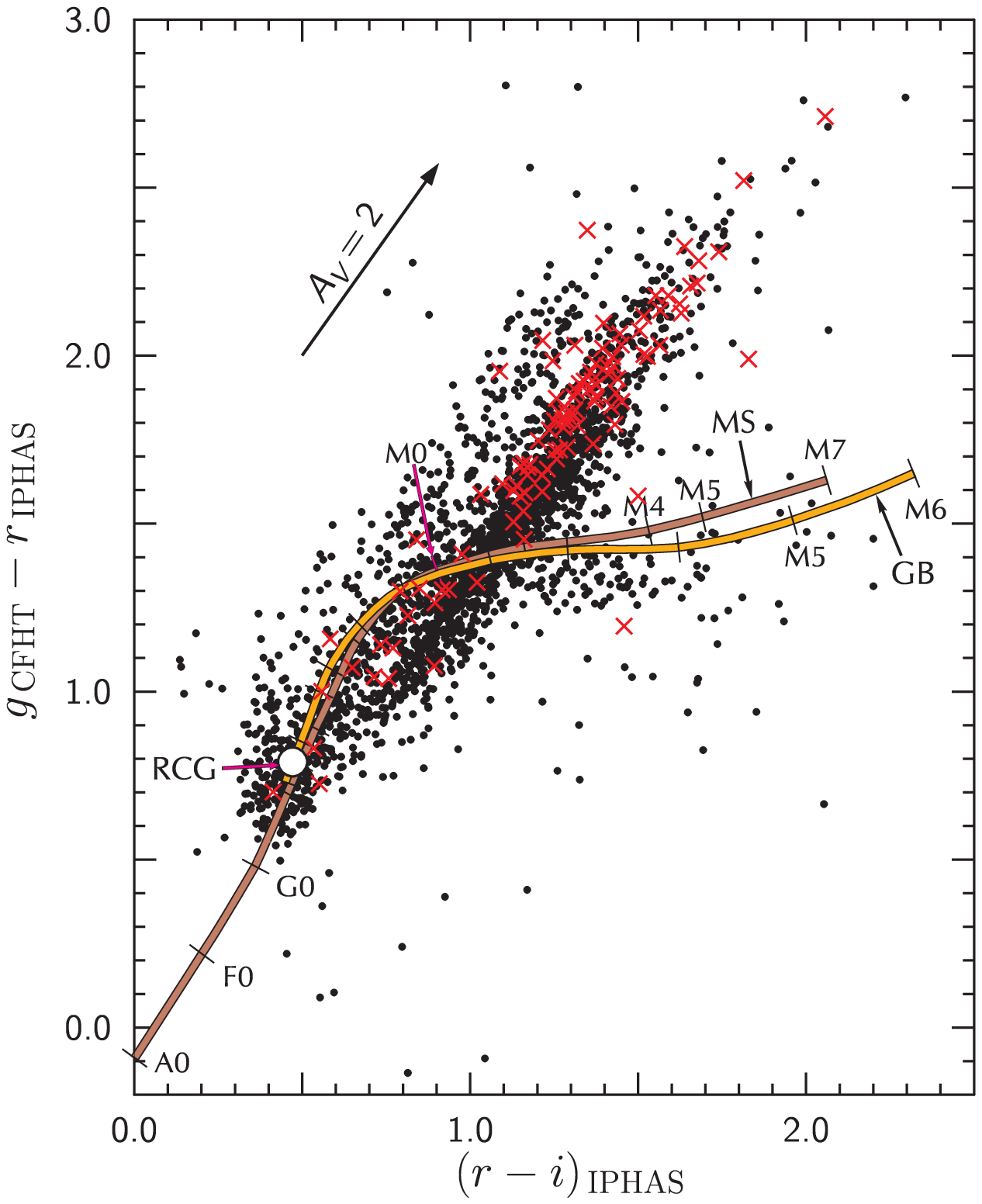,width=100mm,angle=0,clip=}}
\vspace{-.5mm}
\captionb{7}{Observed stars in the $g$--$r$ vs.\,$r$--$i$ diagram.
 The main sequence and the giant branch from Table 1
are shown. Red crosses denote the suspected RCGs.}
}
\end{figure}

The intrinsic sequences of M-type stars, both of dwarfs and giants, turn
sharply to the right since an increase in absorption by TiO bands is
strongly blocking the intensity in the $r$ passband, thus causing
$g$--$r$ to decrease and $r$--$i$ to increase.  However, there is some
disagreement between the values of the observed and synthetic $g$--$r$
colors -- the synthetic sequence is about 0.1 mag above the observed
one.  This can be caused by systematic errors either in spectral energy
distributions or in the response functions of the passbands.  On the
other
hand, the effect of systematic errors in the observed color indices
cannot be excluded.

As in other two-color diagrams, in the $g$--$r$ vs.\,$r$--$i$ diagram
the majority of the identified RCGs (red crosses) form a quite distinct
strip along their reddening lines.  At the same time, a few stars
exhibit considerable deviations from the bulk of points on either side
of the strip.  This means that either their classification as RCGs is
wrong or there are other reasons for such deviations (errors in
photometry, binarity, metal-deficiency, peculiarity, etc.).


\begin{figure}[!th]
\vbox{
\centerline{\psfig{figure=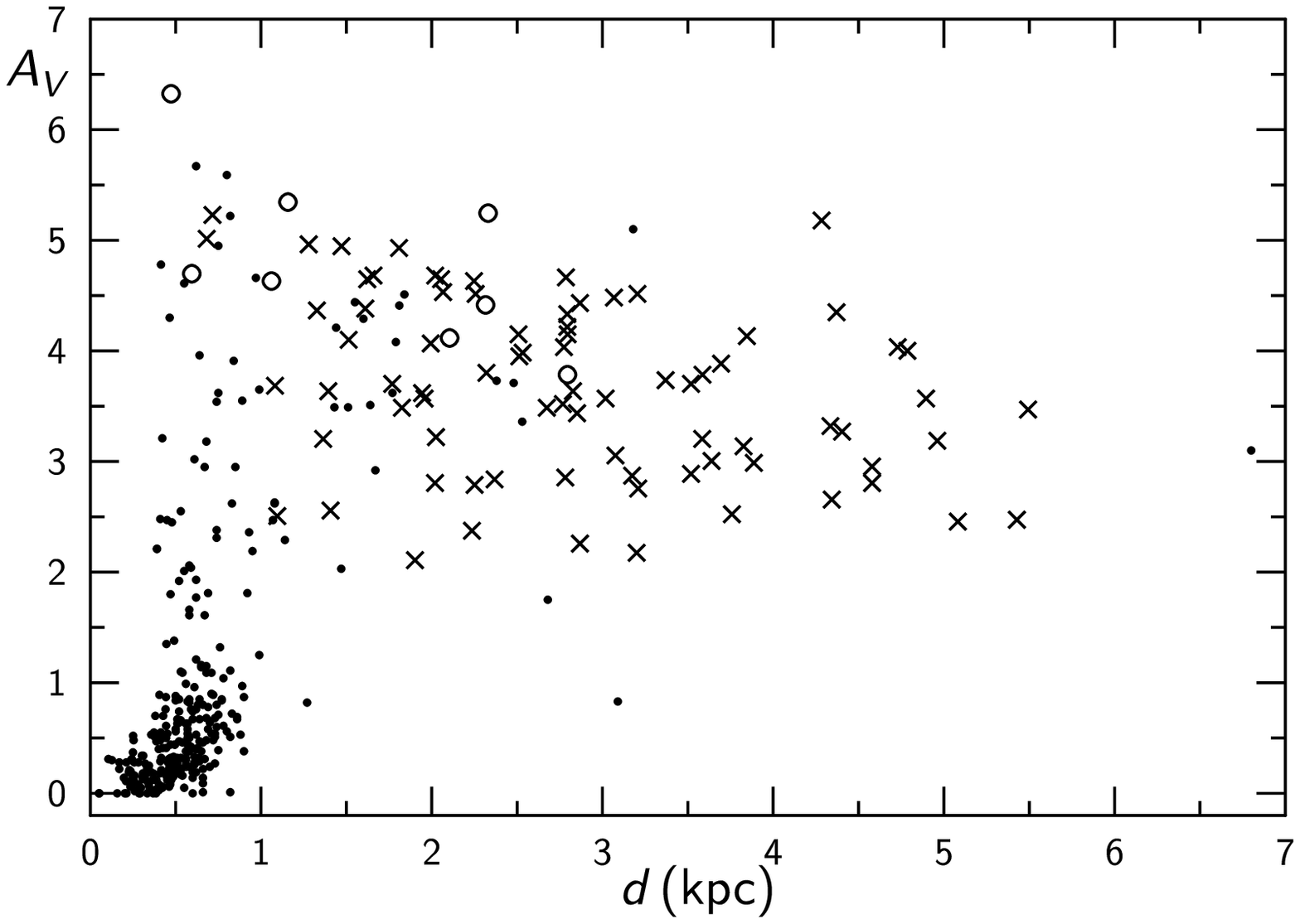,width=124mm,angle=0,clip=}}
\captionb{8}
{Interstellar extinction as a function of the distance up to 7 kpc in
the NAP complex. Dots denote the data for the dark cloud LDN\,935
from Laugalys et al. (2006), open circles denote RCGs in the
dark part of the investigated area, and symbols $\times$ denote RCGs in
the bright part of the area.}}
\end{figure}

\sectionb{9}{THE DEPENDENCE OF EXTINCTION ON DISTANCE}

For stars in Table 2 the values of interstellar
extinction and distance given in the last two columns
(taking $M_{K_s}$ = --1.6) were calculated using the following formulae:
\begin{equation}
A_{K_s} = 2.0\,E_{H-K_s} = 2.0\,[(H-K_s) - 0.09],
\end{equation}
\begin{equation}
5\,\log d = K_s - M_{K_s} +5 - A_{K_s},
\end{equation}
\begin{equation}
A_V = 8.3\,A_{K_s}.
\end{equation}

In Figure 8 the extinctions $A_V$ are plotted against distance, together
with the results obtained in the direction of LDN\,935 by Laugalys et
al.  (2006) from {\it Vilnius} photometry.  The stars with {\it Vilnius}
photometry are plotted as dots, the RCGs from Table 2 in the
direction of the dark cloud as open circles and in the direction of the
apparently transparent parts of the area as crosses.  It is evident that
the values of $A_V$ determined for the newly identified RCGs are in good
agreement with the results from {\it Vilnius} photometry, but extend the
data to greater distances.  The most distant identified RCGs are at
$d$\,$\approx$\,5.5 kpc, i.e., they are located in the Perseus arm.
Their $u$ magnitudes are between 20 and 21, $K_s$ magnitudes are
$\sim$\,12.5 and the $A_V$ extinctions 2.5--3.5 mag.

It should be noted, however, that the RCGs identified using the MegaCam
+ IPHAS data are located mostly in relatively transparent directions, so
they do not give information about the extinction behind of the dust
cloud LDN\,935 which covers a large part of the MegaCam area.  According
to Cambr\'esy et al.  (2002), the extinction $A_V$ in the densest parts
of the cloud can be as large as 35 mag or more.  The limiting magnitudes
in the MegaCam frames are too small to reach RCGs with $A_V$\,$>$\,6
mag.  However, the extinction values up to 5 mag are met even in the
apparently bright areas (for example, in Florida) and at relatively
small distances (0.7--2 kpc).

Seventeen stars of Table 2, not plotted in Figure 8, exhibit the values
of extinction $A_V$ smaller than 2.0 mag at the 2--6 kpc distances.
Other stars at such distances exhibit extinction 2--3 times larger.
These stars are probably misclassified as RCGs, they could infiltrate
through the above conditions from other sequences due to larger
observational errors, peculiarity or duplicity.  Six of such stars are
projected on the dark cloud and are probably G--K dwarfs or subgiants
located in front of the NAP complex.  In Table 2 the numbers of these
stars are marked with asterisks and their extinctions and distances are
not given.

\sectionb{10}{DISCUSSION AND CONCLUSIONS}

Combining deep photometry obtained in the {\it J, H, K}$_s$ system of
the 2MASS survey, the {\it u, g} system of the MegaCam survey and the
{\it r, i} system of the IPHAS survey we proposed a method how to
isolate a group of stars containing the core He-burning red clump giants
with an admixture of shell H-burning giants of spectral classes G5--K1
and M2--M3.  To isolate RCGs the method uses five conditions which
enable to reject stars of other types in the diagrams $J$--$H$
vs.\,$H$--$K_s$, $u$--$g$ vs.\,$r$--$i$ and $K_s$ vs.\,$H$--$K_s$.
Since RCGs are absolutely bright (especially in the near infrared) and
have a tight range of absolute magnitudes and colors, they can be used
for the investigation of interstellar extinction to large distances.

The method was applied to an area of 1\degr\,$\times$\,1\degr\ in the
direction of the North America and Pelican (NAP) nebulae, which covers a
considerable part of the dark cloud LDN\,935 including the Gulf of
Mexico and the `coastal' areas around it.  The identified 87 RCGs, listed
in Table 2, are to be verified in the future by spectral observations.

For the identified RCGs, the values of interstellar extinction and
distances are calculated.  The obtained $A_V$ vs. distance relation in
the area is in agreement with that found by Laugalys et al.  (2006)
using the {\it Vilnius} seven-color system but is extended to larger
distances.  However, for the isolation of RCGs we need an ultraviolet
passband, and this limits the penetration distance.  Fainter stars
should be measured either with longer exposures or with larger
telescopes.

When the ultraviolet magnitudes are not available, we may relax the
conditions and isolate RCGs together with an admixture of red giants
from a wider range of spectral classes -- G5\,III to M3\,III.  In this
case the $J$--$H$ vs.\,$H$--$K_s$ diagram (Figure 2) is sufficient.  The
isolated star sample will be not uniform in absolute magnitudes, but the
difference in the intrinsic $H$--$K_s$ color indices in this range of
spectral classes is only 0.25 mag.  At heavy interstellar reddening such
a dispersion in color excesses is not critical.  Consequently,
background stars on the red giant sequence can be used for mapping the
interstellar extinction of an isolated cloud if their number is
statistically significant.  This method should be more accurate than the
NICE (Lada et al. 1994) and NICER (Lombardi \& Alves 2001) methods which
use stars of all spectral types mixed together.  The main shortcoming of
our method is that it requires a statistically significant number of
background giants observed behind the cloud.

\thanks{ The use of the 2MASS, MegaPipe, IPHAS, SkyView, Gator and
Simbad databases is acknowledged.  This research used the facilities of
the Canadian Astronomy Data Centre operated by the National Research
Council of Canada with the support of the Canadian Space Agency.  We are
thankful to Edmundas Mei\v{s}tas, A.\,G.  Davis Philip and Stanislava
Barta\v{s}i\={u}t\.e for their help in preparing the paper.  Credit for
color picture of the North America and Pelican nebulae:  Adam
Block/NOAO/AURA/NSF.}

\vskip3mm

\References

\refb Alves D. R. 2000, ApJ, 539, 732

\refb Cambr\'esy L., Beichman C. A., Jarrett T. H., Cutri R. M. 2002,
AJ, 123, 2559

\refb Comer\'on F., Pasquali A. 2005, A\&A, 430, 541

\refb Corbally C. J., Strai\v{z}ys V., Laugalys V. 2009, Baltic
Astronomy, 18, 111 (this issue)

\refb Covey K. R., Ivezi\'c \v{Z}., Schlegel D. et al. 2007, AJ, 134,
2398

\refb Dobashi K., Uehara H., Kandori R., Sakurai T., Kaiden M.,
Umemoto T., Sato F. 2005, PASJ, 57, S1

\refb Drew J. E., Greimel R., Irwin M. J. et al. 2005, MNRAS, 362, 753

\refb Grocholski A. J., Sarajedini A. 2002, AJ, 123, 1603

\refb Gwyn S.\,D.\,J. 2008, PASP, 120, 212     

\refb Lada C. J., Lada E. A., Clemens D. P., Bally J. 1994, ApJ, 429,
694

\refb Laugalys V., Strai\v{z}ys V., Vrba F. J., Boyle R. P., Philip
A.\,G.\,D., Kazlauskas A. 2006, Baltic Astronomy, 15, 483

\refb Lombardi M., Alves J. 2001, A\&A, 377, 1023

\refb Lynds B. T. 1962, ApJS, 7, 1

\refb MegaPipe 2009,
http://www2.cadc-ccda.hia-iha.nrc-cnrc.gc.ca/megapipe

\refb Oke J. B. 1965, ARA\&A, 3, 23

\refb Perryman M.\,A.\,C., Lindegren L., Kovalevsky J. et al. 1995,
A\&A, 304, 69

\refb Perryman M.\,A.\,C., Lindegren L., Kovalevsky J. et al. 1997,
A\&A, 323, L49

\refb Strai\v{z}ys V. 1992, {\it Multicolor Stellar Photometry}, Pachart
Publishing House,\\ Tucson, Arizona; available in electronic form from
the author

\refb Strai\v{z}ys V., Laugalys V. 2007, Baltic Astronomy, 16, 327

\refb Strai\v{z}ys V., Laugalys V. 2008a, Baltic Astronomy, 17, 1

\refb Strai\v{z}ys V., Laugalys V. 2008b, Baltic Astronomy, 17, 143

\refb Strai\v{z}ys V., Laugalys V. 2008c, Baltic Astronomy, 17, 253

\refb Strai\v{z}ys V., Corbally C. J., Laugalys V. 2008, Baltic
Astronomy, 17, 125

\refb Strai\v{z}ys V., Lazauskait\.e R. 2009, Baltic Astronomy,
18, 19

\refb Sviderskien\.e Z. 1988, Bull. Vilnius Obs., No.\,80, 3

\refb Sviderskien\.e Z. 1992, Bull. Vilnius Obs., No. 86, 3

\end{document}